\documentclass[12pt,a4paper]{article}
\usepackage{amsmath}
\usepackage{amssymb}
\usepackage{amsthm}
\numberwithin{equation}{section}

\newcommand{\pa}{\partial}

\newcommand{\defeq}{\stackrel{\mathrm{def}}{=}}

\newcommand{\IC}{\mathbb{C}}

\newcommand{\IZ}{\mathbb{Z}}

\newcommand{\frakgl}{\mathfrak{gl}}

\newcommand{\frakg}{\mathfrak{g}}
\newcommand{\fraks}{\mathfrak{s}}
\newcommand{\hfrakg}{\widehat{\mathfrak{g}}}

\newcommand{\hG}{\widehat{G}}
\newcommand{\bft}{\textnormal{\mathversion{bold}$t$}}
\newcommand{\valpha}{\vec{\alpha}}
\newcommand{\vbeta}{\vec{\beta}}

\newtheorem{proposition}{Proposition}
\newtheorem{lemma}{Lemma}
\newtheorem{corollary}{Corollary}

\title{The sixth Painlev\'e equation as \\[2mm]
similarity reduction of $\widehat{\mathfrak{gl}}_3$ hierarchy}
\author{
Saburo K{\sc akei}\\
{\normalsize Department of Mathematics, Rikkyo University}\\
{\normalsize Nishi-ikebukuro, Toshima-ku, Tokyo 171-8501, Japan}\\
{\normalsize E-mail: kakei@rkmath.rikkyo.ac.jp}\\[5mm]%
Tetsuya K{\sc ikuchi}\\
{\normalsize Mathematical Institute, Tohoku University}\\
{\normalsize Aoba-ku, Sendai 980-8578, Japan}\\
{\normalsize E-mail: tkikuchi@math.tohoku.ac.jp}}

\date{}

\begin{document}
\maketitle
\begin{abstract}
Scaling symmetry of 
$\widehat{\mathfrak{gl}}_n$-type 
Drinfel'd-Sokolov hierarchy is investigated. 
Applying similarity reduction to the hierarchy, 
one can obtain the Schlesinger equation
with $(n+1)$ regular singularities. 
Especially in the case of $n=3$, 
the hierarchy contains 
the three-wave resonant system and 
the similarity reduction gives 
the generic case of the Painlev\'e VI equation.
We also discuss Weyl group symmetry 
of the hierarchy.
\end{abstract}

\renewcommand{\thefootnote}{\fnsymbol{footnote}}

\newpage
\section{Introduction}
There are close connections
between soliton equations and the Painlev\'e equations.
For instance, self-similar solutions of the
modified KdV equation satisfy the second 
Painlev\'e equation, and that of the
modified Boussinesq equation satisfy
the fourth Painlev\'e equation.
In these examples, 
soliton equations can be regarded as 
special cases of the $A_l^{(1)}$-type 
Drinfel'd-Sokolov hierarchies.
Along this line,
Noumi and Yamada obtained 
the $A_l^{(1)}$ Painlev\'e system 
\cite{NYA,ICM} as a similarity reduction of 
the Drinfel'd-Sokolov hierarchy of the 
principal $A_l^{(1)}$-type.

In recent works \cite{KK1,KK2}, 
we developed a technique that connects
Lax formalism of soliton equations and 
monodromy preserving deformation, 
based on similarity reduction of the generalized 
Drinfel'd-Sokolov hierarchies. 
As consequences, the Painlev\'e IV equation 
can be obtained from derivative nonlinear 
Schr\"odinger equation \cite{KK1,KK2}, 
and the Painlev\'e V equation from the modified 
Yajima-Oikawa equation \cite{KIK}.

In this paper, extending these works,
we will show a connection between 
the three-wave resonant system,
\begin{equation}
\left\{\begin{aligned}
\partial_\tau u_1+c_1\partial_x u_1 &= i\gamma_1 u_2^*u_3^*,\\
\partial_\tau u_2+c_2\partial_x u_2 &= i\gamma_2 u_3^*u_1^*,\\
\partial_\tau u_3+c_3\partial_x u_3 &= i\gamma_3 u_1^*u_2^*,
\end{aligned}\right.
\label{3Weq}
\end{equation}
and the Painlev\'e VI equation, 
\begin{align}
 \frac{d^2y}{dt^2} = &
\frac{1}{2}\left( 
\frac{1}{y} + \frac{1}{y-1} + \frac{1}{y-t} 
\right)\left(\frac{dy}{dt}\right)^2
-\left(\frac{1}{t} + \frac{1}{t-1} + \frac{1}{y-t} 
 \right)\frac{dy}{dt}
\nonumber\\
& \quad + \frac{y(y-1)(y-t)}{t^2(t-1)^2}
\left\{ \alpha + \beta \frac{t}{y^2}
+ \gamma \frac{t-1}{(y-1)^2}
+ \delta \frac{t(t-1)}{(y-t)^2} \right\},
\label{P6}
\end{align}
where $\alpha$, $\beta$, $\gamma$, $\delta$ are complex 
parameters.

The relation between the three-wave resonant system \eqref{3Weq} 
and the Painlev\'e VI equation \eqref{P6} 
has been studied by considering self-similar solutions 
of \eqref{3Weq}
\cite{FY,Kit}.
Fokas and Yortsos \cite{FY} obtained one-parameter family
of Painlev\'e VI transcendents. 
Based on this work, Kitaev obtained a two-parameter family \cite{Kit}.
In this paper, we introduce a hierarchy of soliton equations 
that includes the three-wave resonant system and 
show that the generic case of the Painlev\'e VI equation 
with four parameters can be obtained as 
similarity reduction of the hierarchy.

Our method to obtain the relation between 
\eqref{3Weq} and \eqref{P6} is based on 
a transformation that maps the Lax equations of 
the three-wave resonant system, whose coefficients
are $3 \times 3$ matrices, 
into the Schlesinger system of the form, 
\begin{equation}
\label{2by2Fuchsian}
\left\{
\begin{aligned}
\frac{\partial Y(x)}{\partial x} &=
\mathcal{A}(x)Y(x), 
\quad
\mathcal{A}(x)
\defeq \frac{\mathcal{A}_0}{x}
+\frac{\mathcal{A}_1}{x-1}
+\frac{\mathcal{A}_t}{x-t},\\
\frac{\partial Y(x)}{\partial t} &=
-\frac{\mathcal{A}_t}{x-t} Y(x),
\end{aligned}\right.
\end{equation}
where $A_i$ ($i=0,1,t$) are $2 \times 2$ matrices 
independent of $x$ and obey the following conditions:
\begin{equation}
\begin{aligned}
&\det\mathcal{A}_i=0, \qquad
\mbox{eigenvalues of }\mathcal{A}_i\mbox{ are } 0
 \mbox{ and }  \theta_i
\quad (i=0,1,t),\\
&-\mathcal{A}_0-\mathcal{A}_1-\mathcal{A}_t=
\begin{pmatrix}
\kappa_1 & 0\\ 0 & \kappa_2
\end{pmatrix}.
\end{aligned}
\end{equation}
It is well-known that the  
Painlev\'e VI equation is obtained from the 
Schlesinger system \eqref{2by2Fuchsian} \cite{JM2}
(cf. \cite{Ok}). 
The zero of the $(1,2)$-element of the coefficient 
matrix $\mathcal{A}(x)$, i.e., 
\begin{equation}
\label{solOfP6}
y(t) = \frac{-t(\mathcal{A}_1)_{12}}
     {(\mathcal{A}_2)_{12} 
     +t(\mathcal{A}_3)_{12}}, 
\end{equation}
solves the Painlev\'e VI equation \eqref{P6}
with the parameters,
\begin{equation}
 \alpha 
= \frac{(\kappa_1 - \kappa_2 - 1)^2}{2},
\quad
 \beta = -\frac{\theta_0^2}{2},
\quad
 \gamma = \frac{\theta_1^2}{2},
\quad
 \delta = \frac{1-\theta_t^2}{2}.
\end{equation}

Harnad \cite{Har} and Mazzocco \cite{Maz} 
gave the description of the Painlev\'e VI
in terms of $3 \times 3$ linear system with 
a simple pole at $0$ and a double pole at $\infty$; 
\begin{equation}
\label{33Lax}
z\frac{d\Psi}{dz}
= (T+V)\Psi,
\qquad 
 T = \mathrm{diag}(0, t, 1)
\end{equation}
They discussed the transformation of the 
$3 \times 3$ system \eqref{33Lax} 
into the $2 \times 2$ system \eqref{2by2Fuchsian}.
In this paper, 
we will show that the $3\times 3$ linear problem \eqref{33Lax} 
and its Laplace transformation to \eqref{2by2Fuchsian} can be 
reformulated as similarity reduction of the 
Drinfel'd-Sokolov hierarchy of $\widehat{\mathfrak{gl}}_3$-type.

Note that if the matrix $V$ in \eqref{33Lax} is 
skew symmetric, 
the soliton equation associated with \eqref{33Lax} 
is equivalent to the Darboux-Egoroff equation, which is a 
special case of \eqref{3Weq} and characterize the metric 
behind the solution of the Witten-Dijkgraaf-Verlinde-Verlinde 
equations \cite{Dub,AV}. 
As discussed in \cite{Dub,AV},
the deformation equations of \eqref{33Lax} with 
skew symmetric $V$ gives 
a one-parameter family $(\alpha = -\beta, \gamma=0, 
\delta = 1/2)$ of the Painlev\'e VI transcendents, 
whereas we do not impose the skew-symmetric condition for $V$ 
in this paper.
Therefore we obtain the generic case of the Painlev\'e VI.

This paper is organized as follows.
In section \ref{section2}, we review our construction of
the soliton equations by using the affine Lie algebra $\widehat{\frakgl}_n$.
The three-wave resonant system \eqref{3Weq} can be treated as 
the case $n=3$. 
In section \ref{section3}, we discuss scaling symmetry of 
the $\widehat{\frakgl}_n$ hierarchy, and consider similarity 
reduction of the hierarchy. 
If we consider the $\widehat{\frakgl}_3$ case, we can 
obtain the Painlev\'e VI equation with full parameters. 
Weyl group symmetry of the hierarchy is 
discussed in Section \ref{sec:WeylGp}.
Section \ref{sec:ConcludingRemarks} is devoted to 
concluding remarks.

\section{$\widehat{\mathfrak{gl}}_n$ hierarchy 
and the three-wave resonant system}
\label{section2}

In this section we outline our formulation of soliton 
equations \cite{KK1}, which is 
based on the affine Lie group approach of 
Drinfel'd and Sokolov \cite{DS} and its 
generalizations \cite{BtK2, gds1, Wil}. 
We remark that we use only the homogeneous case 
in this paper, to treat 
the three-wave resonant system and the Painlev\'e VI
equation.

Let $\frakg = \frakgl_n(\IC)$.
The affine Lie algebra $\widehat{\frakg} 
= \widehat{\mathfrak{gl}}_n$ is
realized as a central extension of the loop
algebra $\frakg\otimes\IC[z,z^{-1}]$ together with 
the derivation $d=d/dz$:
\begin{equation}
 \hfrakg = \frakg\otimes\IC[z,z^{-1}]\oplus\IC K\oplus\IC d.
\end{equation}
The Lie bracket is given as follows:
\begin{equation}
\begin{aligned}
 &[X \otimes z^j, Y\otimes z^k] =
  [X,Y] \otimes z^{j+k}+ j \delta_{j+k,0}(X,Y)K, 
\\
 &[K,\hfrakg]=0,
\qquad 
 [d, X\otimes z^j] = jX\otimes z^j, 
\end{aligned}
\end{equation}
for $X,Y\in \mathfrak{gl}_n$, 
$j, k \in\IZ$ and $(X, Y)$ is 
the normalized invariant scalar product 
of $\frakg$ \cite{Kac}. 
The derivation $d$ induces a $\IZ$-grading on $\hfrakg$, which
is called the homogeneous gradation: 
\begin{equation}
 \hfrakg = \mathop{\bigoplus}_{j\in\IZ}\hfrakg_j, \quad 
 \hfrakg_j =\{x\in\hfrakg\; | \;[d, x]=jx\}.
\end{equation}
For an integer $k$, we use the notation 
\begin{equation}
 \hfrakg_{\ge k} = 
\mathop{\bigoplus}_{j\ge k}\hfrakg_j, \quad
 \hfrakg_{<k} = 
\mathop{\bigoplus}_{j<k}\hfrakg_j.
\end{equation}

To construct integrable hierarchies, 
Heisenberg subalgebras of 
$\widehat{\mathfrak{gl}}_n$ play a crucial role. 
In general, nonequivalent Heisenberg subalgebras 
are classified by conjugacy classes of 
the Weyl group of $\frakg$ \cite{gds1}. 
In the case of $A_l^{(1)}$, conjugacy classes are 
labeled by partitions $\{(n_1, \dots, n_s);
n_1 + \cdots + n_s = n\}$. 
In this paper, we consider the homogeneous Heisenberg 
subalgebra $\widehat{\fraks}$ that is associated with the 
partition $(1,1,\dots, 1)$.
In this case, we can choose a graded
basis of $\widehat{\fraks}$ as 
\begin{equation}
 \Lambda_k^a \defeq z^k E_{aa}
\qquad
 (1 \le a \le n, \;
k \in \IZ \setminus \{ 0 \})
\end{equation}
where $E_{aa} = (\delta_{ia}\delta_{ja})$ is the matrix unit.
These basis of $\widehat{\fraks}$ are
related to the homogeneous gradation:
\begin{equation}
 [d, \Lambda_k^a] = k\Lambda_k^a. 
\end{equation}

We now consider a Kac-Moody group $\hG$ formed by exponentiating 
the action of $\hfrakg$ on a integrable module. 
Throughout this paper, we assume that the exponentiated action of 
an element of the positive degree subalgebra
of $\widehat{\fraks}$ is well-defined. 
We remark that all of the representations used in what follows 
belong to this category.
We denote by $\hG_{\ge 0}$ and 
$\hG_{<0}$ the subgroups of $\hG$ correspond 
to the subalgebras $\hfrakg_{\ge 0}$ and $\hfrakg_{<0}$, 
respectively. 

We construct a hierarchy of soliton equations
with time valuables $\bft =(t_j^1, \dots , t_j^n)_{j > 0}$.
First we define 
\begin{equation}
\label{wave}
 \Psi_0(z; \bft) 
\defeq \exp \Bigl( \sum_{j>0} \sum_{a=1}^n 
  t_j^a \Lambda_j^a \Bigr)
= \exp \Bigl( \sum_{a=1}^n \Bigl( \sum_{j > 0} 
  z^j t_j^a \Bigr) E_{aa} \Bigr),
\end{equation}
Starting from an element $g(z;0) \in \hG$, 
we define the time-evolution by
\begin{equation}
\label{defg(t)}
 g(z;\bft)\defeq 
 \Psi_0(z;\bft) g(z;0).
\end{equation}
Then $g(z; \bft)$ satisfies 
the following differential equations, 
\begin{equation}
\label{LinearDE}
 \frac{\pa g(z;\bft)}{\pa t_j^a}=\Lambda_j^a g(z;\bft) 
\qquad 
(1 \le a \le n, \; j > 0).
\end{equation}
In what follows, we shall assume the existence
of the Gauss decomposition
with respect to the homogeneous gradation: 
\begin{align}
g(z; \bft) &= \{g_{<0}(z; \bft)\}^{-1}g_{\ge 0}(z; \bft), 
\label{GaussDecomp}\\
g_{<0}(z; \bft) &= 1 + g_{-1} + g_{-2} + \cdots  
\in \hG_{<0}, \label{g<0}
\\
g_{\ge 0}(z; \bft) &=  g_0(1 + g_1 + g_2 + \cdots)
\in \hG_{\ge 0}. \label{g>=0} 
\end{align}
A detailed discussion about this assumption is 
in \cite{BtK1,Wil} for instance. 
\begin{lemma}
\label{lemma:uniqueness}
Suppose that $g\in\widehat{G}$ is invertible and 
the Gauss decomposition \eqref{GaussDecomp} is 
achieved. Then the decomposition is unique.
\end{lemma}
A proof can be found in \cite{KK1}, which is 
an analogue of the theorem 4.1 in \cite{UT} that 
treats the category of infinite matrices.

{}From \eqref{LinearDE} and \eqref{GaussDecomp}, 
we have the following equations
for $g_{<0} = g_{<0}(z;\bft)$ and 
$g_{\ge 0} = g_{\ge 0}(z;\bft)$: 
\begin{align}
\frac{\pa g_{<0}}{\pa t_j^a} &
= -\bigl(g_{<0} \Lambda_j^a g_{<0}^{-1} \bigr)_{<0}g_{<0}
=  B_j^a g_{<0}- g_{<0}\Lambda_j^a, 
\label{SatoEq<} 
\\
\frac{\pa g_{\ge 0}}{\pa t_j^a} &= B_j^a g_{\ge 0}
\quad (1 \le a \le n,\; j >0),
\label{SatoEq>=}
\end{align}
where $B_j^a = B_j^a(z;t)$ is defined by 
\begin{equation}
  B_j^a \defeq 
 \left(g_{<0} \Lambda_j^a g_{<0}^{-1} \right)_{\ge 0}
\in \hfrakg_{\ge 0}.
\label{defofBn}
\end{equation}
We call \eqref{SatoEq<} and \eqref{SatoEq>=} 
the Sato-Wilson equations.
Note that the equation \eqref{SatoEq<} imply 
the condition: 
\begin{equation}
\partial g_{<0} 
\defeq \sum_{a=1}^n \frac{\partial g_{<0}}{\partial t_1^a}
= 0, 
\label{traceless}
\end{equation}
and hence the hierarchy of equations \eqref{SatoEq<} is
equivalent to the $(1, \dots, 1)$-reduction of 
the $n$-component KP hierarchy \cite{UT}.

The compatibility conditions for \eqref{SatoEq<} or \eqref{SatoEq>=} 
give rise to the Zakharov-Shabat (or zero-curvature) equations, 
\begin{equation}
\label{ZakShab}
\left[ \frac{\pa}{\pa t_j^a} - B_j^a,\;
       \frac{\pa}{\pa t_k^b} - B_k^b \right] =0
\qquad 
 (1 \le a,b \le n, \quad
      j, k > 0)
\end{equation}
The family of equations of \eqref{ZakShab} 
gives a hierarchy of soliton equations.

We define the Baker-Akhiezer functions with 
complex parameters $\vec{\alpha} 
= (\alpha_1, \dots, \alpha_n)$ and $\vec{\beta} 
= (\beta_1, \dots, \beta_n)$ by
\begin{align}
 &\Psi^{(\infty)}(z; \bft, \vec{\alpha})
\defeq g_{<0}(z; \bft) \cdot \Psi_0(z; \bft) \cdot z^{D(\vec{\alpha})},
\label{Psi-infty} \\
 &\Psi^{(0)}(z;\bft, \vec{\beta})
\defeq g_{\ge 0}(z; \bft) \cdot z^{D(\vec{\beta})}.
\label{Psi-zero}
\end{align}
Here we have set $D(\vec{\alpha}) 
= \mathrm{diag}(\alpha_1, \dots, \alpha_n)$ and $D(\vec{\beta})
= \mathrm{diag}(\beta_1, \dots, \beta_n)$.
Both of the functions $\Psi^{(\infty)}$ and
$\Psi^{(0)}$ satisfy the following equations
\begin{equation}
\partial \Psi
\defeq \sum_{a=1}^n
 \frac{\partial \Psi}{\partial t_1^a}
= z\Psi,\\
\qquad
 \frac{\partial \Psi}{\partial t_j^a}
= B_j^a \Psi
\qquad (1 \le a \le n,\; j > 0)
\end{equation}
by definitions and the Sato-Wilson equations 
\eqref{SatoEq<} and \eqref{SatoEq>=}.

To obtain the three-wave resonant system \eqref{3Weq}, 
we now restrict ourselves to 
the $\widehat{\mathfrak{gl}}_3$ case.
We introduce the following parameterizations for the 
matrices $g_{-1}$ of \eqref{g<0} and $g_0$ of \eqref{g>=0}:
\begin{equation}
g_{-1} = 
\begin{pmatrix}
 w_{11} & w_{12} & w_{13} \\
 w_{21} & w_{22} & w_{23} \\
 w_{31} & w_{32} & w_{33} 
\end{pmatrix}
z^{-1},
\qquad
g_0 = 
\begin{pmatrix}
 g_{11} & g_{12} & g_{13} \\
 g_{21} & g_{22} & g_{23} \\
 g_{31} & g_{32} & g_{33} 
\end{pmatrix}.
\label{Sato-WilsonOp}
\end{equation}
Using these parameterizations, we can 
express $B_1^a$ $(a=1,2,3)$ as follows:
\begin{align}
 B_1^1 &= 
\begin{pmatrix}
 z   & -w_{12} & -w_{13} \\
 w_{21} &     0   &    0    \\
 w_{31} &     0   &    0 
\end{pmatrix},\\
 B_1^2  &= 
\begin{pmatrix}
     0   &  w_{12}  &    0 \\
 -w_{21} &    z     & -w_{23} \\
     0   &  w_{32}  &    0 
\end{pmatrix},
\\
 B_1^3 
     &= 
\begin{pmatrix}
     0   &    0     & w_{13} \\
     0   &    0     & w_{23} \\
 -w_{31} & -w_{32}  &    z
\end{pmatrix}.
\end{align}
The Zakharov-Shabat equation \eqref{ZakShab} gives 
the following equations for $w_{ij}$ $(i \neq j)$:
\begin{align}
 &\frac{\partial w_{23}}{\partial t_1^1}
= w_{21}w_{13},
\qquad
 \frac{\partial w_{32}}{\partial t_1^1}
= w_{31}w_{12},
\label{3KP1} \\
 &\frac{\partial w_{13}}{\partial t_1^2}
= w_{12}w_{23},
\qquad
 \frac{\partial w_{31}}{\partial t_1^2}
= w_{32}w_{21},
\label{3KP2} \\
 &\frac{\partial w_{12}}{\partial t_1^3}
= w_{13}w_{32},
\qquad
 \frac{\partial w_{21}}{\partial t_1^3}
= w_{23}w_{31}.
\label{3KP3}
\end{align}
Moreover, $w_{ij}$ satisfy
\begin{equation}
 \partial w_{ij} = 0,
\qquad \partial =
\frac{\partial}{\partial t_1^1}
+ \frac{\partial}{\partial t_1^2}
+ \frac{\partial}{\partial t_1^3},
\label{const}
\end{equation}
because of \eqref{traceless}. 
We remark that this condition corresponds to the 
$(1,1,1)$-reduction of the $3$-component KP hierarchy 
and has been discussed in \cite{KvdL}. 
We restrict the time variables to the plane, 
\begin{equation}
 t_1^k = a_k x + b_k \tau
\qquad
 (k=1,2,3),
\label{4.1.5}
\end{equation}
where $a_k$, $b_k$ are complex parameters. 
Under these conditions, 
we arrive at the three-wave resonant system \eqref{3Weq} 
by setting $u_1=w_{23}$, $u_1^*=w_{32}$, 
$u_2=w_{31}$, $u_2^*=w_{13}$, 
$u_3=w_{12}$, $u_3^*=w_{21}$, and applying suitable scaling.

\section{Similarity reduction to Schlesinger system}
\label{section3}
\subsection{Extended scaling symmetry and similarity reduction}
\label{subsec:ExtendedScalingSymmetry}
In this subsection, we consider scaling symmetry of 
the $\widehat{\frakgl}_n$ homogeneous hierarchy.
For $\vec{\alpha}=(\alpha_1, \dots, \alpha_n)$, $\vec{\beta}
= (\beta_1, \dots, \beta_n) \in\IC^n$ and 
$\lambda\in\IC^{\times}$, we define 
$\tilde{g}_{<0}(z; \bft)$, $\tilde{g}_{\ge 0}(z; \bft)$ and 
$\bft_\lambda$ as follows: 
\begin{align}
&\tilde{g}_{<0}(z; \bft) \defeq  
\lambda^{D(\vec{\alpha})}
g_{<0}(\lambda^{-1}z; \bft_\lambda)
\lambda^{-D(\vec{\alpha})},
\label{g<0tilde}
\\
&\tilde{g}_{\ge 0}(z; \bft) \defeq  
\lambda^{D(\vec{\alpha})}
g_{\ge 0}(\lambda^{-1}z; \bft_\lambda)
\lambda^{-D(\vec{\beta})}
\label{g>0tilde},\\
&\bft_\lambda 
\defeq (\lambda^j t_j^1, 
 \dots, \lambda^j t_j^n)_{j > 0}.
\end{align}
We remark that one can set $\beta_n=-1$ without loss of 
generality. This choice of $\beta_n$ is important 
to obtain the Schlesinger system.
\begin{proposition}
\label{prop1}
If $g_{<0}$ and $g_{\ge 0}$ solve 
the Sato-Wilson equations \eqref{SatoEq<} and \eqref{SatoEq>=}, 
then 
$\tilde{g}_{<0}(z; \bft)$ and $\tilde{g}_{\ge 0}(z; \bft)$
also satisfy \eqref{SatoEq<} and \eqref{SatoEq>=}.
\end{proposition}
\begin{proof}
Because of the relation
$\Psi_0(\lambda^{-1}z; \bft_\lambda) 
= \Psi_0(z; \bft)$ and 
$[\lambda^{D(\vec{\alpha})}, \Lambda_j^a] = 0$, we find that
\begin{equation}
\label{transformedg}
 \lambda^{D(\vec{\alpha})} g(\lambda^{-1}z, \bft_\lambda)
 \lambda^{-D(\vec{\beta})}
= \Psi_0(z; \bft) \lambda^{D(\vec{\alpha})} g(\lambda^{-1}z,0)
 \lambda^{-D(\vec{\beta})}.
\end{equation}
Therefore the function \eqref{transformedg}
satisfies the differential equation \eqref{LinearDE}.
Due to the uniqueness of the the Gauss decomposition,
we have the desired result.
\end{proof}

Proposition \ref{prop1} is an extension of
the Virasoro symmetry of the generalized 
Drinfel'd-Sokolov hierarchy \cite{BdGHM}.
In the case of $n = 3$, 
the parameters $\vec{\alpha}$ and $\vec{\beta}$ correspond to 
those of the Painlev\'e VI (See \eqref{P6para} below).

We now impose a constraint to the 
initial data $g(z;0)$ as 
\begin{equation}
\label{Similarity_to_g(z;0)}
  g(z;0) = \lambda^{D(\vec{\alpha})} 
 g(\lambda^{-1}z;0) \lambda^{-D(\vec{\beta})}.
\end{equation}
This constrained initial data gives the following self-similar 
solutions of the equations \eqref{LinearDE},
\eqref{SatoEq<} and \eqref{SatoEq>=}:
\begin{gather}
\label{simg}
 g(z; \bft) 
= \lambda^{D(\vec{\alpha})}  
 g(\lambda^{-1}z; \bft_\lambda) 
  \lambda^{-D(\vec{\beta})}
\\
\label{simg<}
 g_{<0}(z; \bft) 
= \lambda^{D(\vec{\alpha})} 
 g_{<0}(\lambda^{-1}z; \bft_\lambda)
  \lambda^{-D(\vec{\alpha})},
\\
\label{simg>=}
 g_{\ge 0}(z; \bft) 
= \lambda^{D(\vec{\alpha})} 
 g_{\ge 0}(\lambda^{-1}z; \bft_\lambda)
  \lambda^{-D(\vec{\beta})}.
\end{gather}
\begin{proposition}
\label{prop2}
Under the similarity condition \eqref{Similarity_to_g(z;0)},
the Baker-Akhiezer functions 
$\Psi^{(\infty)}(z; \bft, \vec{\alpha})$ and 
$\Psi^{(0)}(z; \bft, \vec{\beta})$ satisfy the 
linear differential equations, 
\begin{align}
 &z\frac{\partial \Psi}{\partial z} = 
\Bigl( D(\valpha) + \sum_{a=1}^n \sum_{j>0} jt_j^a B_j^a \Bigr)\Psi,
\label{senkei}
\\
 &\frac{\partial \Psi}{\partial t_j^a} = B_j^a \Psi
\quad
 (1 \le a \le n, \; j > 0)
\label{henkei}
\end{align} 
\end{proposition}
\begin{proof}
Differentiating \eqref{simg<} 
with respect to $\lambda$ at $\lambda = 1$, 
we obtain the equations
\begin{align}
 &z\frac{\partial g_{<0}}{\partial z} 
= [D(\valpha), g_{<0} ] 
+ \sum_{a=1}^n \sum_{j>0} jt_j^a 
\frac{\partial g_{<0}}{\partial t_j^a},
\label{simW}
\\
 &z\frac{\partial g_{\ge 0}}{\partial z} 
= D(\valpha) g_{\ge 0}
- g_{\ge 0} D(\vbeta)
+ \sum_{a=1}^n \sum_{j>0} jt_j^a 
\frac{\partial g_{\ge 0}}{\partial t_j^a}.
\label{simZ}
\end{align}
{}From the definition of the
Baker-Akhiezer functions $\Psi^{(\infty)}$ and $\Psi^{(0)}$, 
we have \eqref{senkei}, \eqref{henkei}.
\end{proof}
Note especially that we can obtain the following relation 
from the grade $0$ part of \eqref{simZ}: 
\begin{equation}
\label{diagonalization}
 g_0^{-1} \Bigl( D(\valpha) 
+ \sum_{a=1}^n \sum_{j>0} jt_j^a B_{j,0}^a  \Bigr)
 g_0 = D(\vbeta).
\end{equation}
Here $B_{j,0}^a$ denotes the grade $0$ part
of $B_j^a$, i.e., 
$B_{j,0}^a=(g_{<0}\Lambda_j^a g_{<0}^{-1})_0$.

\begin{description}
\item[Remark]
In the case of the Darboux-Egoroff equations \cite{AV}, 
initial data $g(z) = g(z;0)$ should satisfy the condition
$g^{-1}(z) = g^T(-z)$. 
It follows that $g_{<0}$ and $g_{\ge 0}$ satisfy
\begin{equation}
 g_{<0}^{-1}(z; \bft) = g_{<0}^T(-z; \bft),
\quad
 g_{\ge 0}^{-1}(z; \bft) = g_{\ge 0}^T(-z; \bft).
\end{equation}
Especially, we have 
\begin{equation}
 g_0^T = g_0^{-1},
\quad
 g_{-1} = g_{-1}^T.
\end{equation}
In addition, 
their formulation of the Virasoro condition does not 
have the parameters that correspond to $\vec{\alpha}$.
\end{description}

\subsection{Transformation to the Schlesinger system}
In the following, we set 
\begin{equation}
  t_i \defeq t_1^i,
\qquad 
  t_j^i = 0 
\quad 
  (1 \le i \le n,\; j>1).
\end{equation}
Then the equation \eqref{senkei} is rewritten as
\begin{equation}
\label{Birkhoff}
 z\frac{\partial \Psi}{\partial z} = (zT + V)\Psi,
\end{equation}
where $T$ and $V$ are $n\times n$ matrices given by 
\begin{equation}
 T \defeq 
 \mathrm{diag}(t_1, \dots, t_n),
\qquad 
 V \defeq 
  D(\vec{\alpha}) + \sum_{a=1}^n t_a B_{1,0}^a.
\end{equation} 
Notice that the diagonal elements of $V$ are given by 
$\alpha_i$ because those of 
$B_{1,0}^i = [g_{-1},\Lambda_1^i]$ vanish. 
{}From the relation \eqref{diagonalization}, we can show that 
$V$ is diagonalized by $g_0$:
\begin{equation}
g_0^{-1} V g_0 = D(\vec{\beta}).
\end{equation}

The linear equation \eqref{Birkhoff} has 
a simple pole at $z=0$ and a double pole at $z=\infty$.
The solution $\Psi^{(\infty)}$ corresponds to
the fundamental matrix at $\infty$ and 
the parameters $\alpha_i$ to the monodromy index at $\infty$.
Another solution $\Psi^{(0)}$ corresponds
to the fundamental matrix at $0$, and
$\beta_i$ to the monodromy index at $0$.

To rewrite the linear problem \eqref{Birkhoff} as 
the Schlesinger system \eqref{2by2Fuchsian}, 
we apply the Laplace transformation to the solution $\Psi(z)$ 
of the linear equation \eqref{Birkhoff} \cite{Har,Maz}:
\begin{equation}
\label{LaplaceT}
 \Psi(z) \mapsto
\Phi(x) = 
L[\Psi(z)](x)
\defeq \int_\gamma e^{-zx}\Psi(z)dz.
\end{equation}
Here the contour $\gamma$ is chosen to satisfy
\begin{equation}
 \int_\gamma \frac{\partial f}{\partial z} dz = 0.
\end{equation}
Under this choice, the transformation \eqref{LaplaceT} 
has the properties, 
\begin{equation}
\left\{
\begin{aligned}
  \left(\frac{\partial}{\partial x}\right)^k L[\Psi(z)](x) 
   &=  L[(-z)^k \Psi(z)],
\\
 \quad x^k L[\Psi(z)](x)
   &= L\Bigl[\left(\frac{\partial}{\partial z} \right)^k \Psi(z)\Bigr].
\end{aligned}
\right.
\end{equation}
By direct calculation, 
we see that the transformed solution $\Phi(x)$ satisfies
the linear equation, 
\begin{equation}
\label{Hyper}
 (xI - T)\frac{\partial \Phi(x)}{\partial x}
= -(V + I)\Phi(x),
\end{equation}
where $I$ is the unit matrix.
Notice that we can rewrite \eqref{Hyper} as 
\begin{equation}
\label{Schlesinger}
\frac{\partial \Phi(x)}{\partial x}
= -\sum_{j=1}^n \frac{E_{jj}(V+I)}{x - t_j}\Phi(x), 
\end{equation}
which is an $n \times n$ Schlesinger system 
with $(n+1)$ regular singularities at $t_1, \dots, t_n, \infty$.

Remember the Baker-Akhiezer function $\Psi^{(0)}(z; \bft, \valpha)$
of \eqref{Psi-zero} has the form, 
\begin{equation}
 \Psi^{(0)}(z; \bft,\valpha) = g_{\ge 0}(z)z^{D(\valpha)}
    = g_0(1 + g_1 + \cdots)z^{D(\valpha)}.
\end{equation}
The normalized solution at $z=0$ is
obtained as $\tilde{\Psi} = g_0^{-1}\Psi(z; \bft)$.
We denote as $\tilde{\Phi}(x; \bft)$ 
the Laplace transformation of $\tilde{\Psi}$:
\begin{equation}
\label{Phitilde}
 \tilde{\Phi}(x; \bft) \defeq  L[\tilde{\Psi}(z; \bft)] 
= g_0^{-1}\Phi(x; \bft).
\end{equation}
We have the following corollary of Proposition \ref{prop2}:
\begin{corollary}
The function $\tilde{\Phi}(z;\bft)$ of $\eqref{Phitilde}$ satisfies 
the system of equations
\begin{align}
\frac{\partial \tilde{\Phi}(z;\bft)}{\partial x}
&= \sum_{i=1}^n
\frac{A_i(\bft)}{x - t_i}
\tilde{\Phi}(z;\bft), \label{66_a}\\
 \frac{\partial \tilde{\Phi}(z;\bft)}{\partial t_i}
&= -\frac{A_i(\bft)}{x - t_i} \tilde{\Phi}(z;\bft)
\quad (1 \le i \le n),\label{66_b}
\end{align}
where we have defined $A_i(\bft)$ $(1 \le i \le n)$ as
\begin{align}
 A_i(\bft) &\defeq -g_0^{-1} E_{ii}(V+I) g_0
     = -g_0^{-1}E_{ii}g_0 (D(\vbeta) + I)
\label{def_Aa}.
\end{align}
\end{corollary}
\begin{proof}
First equation is obtained by \eqref{Schlesinger}
since $g_0$ is independent of $z$.
Now we apply the Laplace transformation to the 
deformation equation \eqref{henkei}.
The linear equation \eqref{henkei} for $j=1$ is reduced to 
\begin{equation}
 \frac{\partial \Psi}{\partial t_i} 
= B_1^i \Psi
= (\Lambda_1^i + [g_{-1},\Lambda_1^i])\Psi.
\end{equation}
The transformed solution $\Phi = L[\Psi]$ satisfies
\begin{equation}
\label{deform1}
 \frac{\partial \Phi}{\partial t_i} 
= -E_{ii} \frac{\partial \Phi}{\partial x} 
+ [g_{-1},\Lambda_1^i]\Phi
= -\frac{E_{ii}(V+I)}{x-t_i}\Phi
+ [g_{-1},\Lambda_1^i]\Phi.
\end{equation}
To obtain the corresponding linear equation for 
$\tilde{\Phi} = g_0^{-1}L[\Psi]$, 
we prepare a differential equation for $g_0$ with
respect to $t_i$:
\begin{equation}
\label{SatoEq0}
 \frac{\partial g_0}{\partial t_i} 
= [g_{-1}, \Lambda_1^i] g_0, 
\end{equation}
which is the grade $0$ part of the Sato-Wilson 
equation \eqref{SatoEq>=}.
Combine \eqref{SatoEq0} and \eqref{deform1}, 
we have the linear equation for $\tilde{\Phi}$:
\begin{align}
 \frac{\partial \tilde{\Phi}}{\partial t_i} 
= -g_0^{-1} \frac{\partial g_0}{\partial t_i} 
   g_0^{-1} + g_0^{-1} \frac{\partial \Phi}{\partial t_i} 
= -\frac{g_0^{-1}E_{aa}(V+I) g_0}{x - t_a}
  \tilde{\Phi}.
\end{align}
Thus we have \eqref{66_b}.
\end{proof}

The system of equations \eqref{66_a} and \eqref{66_b}
give the monodromy preserving deformation of the linear 
equation \eqref{66_a} with respect to 
$t_i$ ($1 \le i \le n$).
Each of $A_i$ satisfies the condition
\begin{equation}
\label{trace_EaaV}
\mathrm{trace}\, A_i = -\alpha_i - 1,
\quad \det A_i = 0
\quad (1 \le i \le n).
\end{equation}
Moreover the relation \eqref{diagonalization} is reduced to 
\begin{equation}
\sum_{i=1}^n A_i = -g_0^{-1}(V + I)g_0 
= -D(\vbeta) - I.
\end{equation}
Under these conditions, the differential equations
\eqref{66_a} and \eqref{66_b} 
can be regarded as the Schlesinger system normalized at $x=\infty$.

Hereafter in this section, we set $n=3$ and 
focus on the $\widehat{\mathfrak{gl}}_3$-case. 
As mentioned in Section \ref{subsec:ExtendedScalingSymmetry}, 
we can set $\beta_3=-1$ without loss of generality.
With this choice, the coefficients $A_i(\bft)$ ($i=1,2,3$) 
have the property $(A_i(\bft))_{j3}= 0$ ($i,j=1,2,3$). 
Thus the two dimensional subspace $\{\,^t(y_1,y_2,0)\,\}$ is 
invariant under the action of $A_i(\bft)$ ($i=1,2,3$) 
with $\beta_3=-1$.
Projecting the equations \eqref{66_a}, \eqref{66_b} to the 
two dimensional subspace, 
we obtain the $2\times 2$ Schlesinger system of the form,
\begin{align}
\frac{\partial Y(x;\bft)}{\partial x}
&= \left\{
 \frac{\tilde{A}_1(\bft)}{x - t_1}
 + \frac{\tilde{A}_2(\bft)}{x - t_2}
 + \frac{\tilde{A}_3(\bft)}{x - t_3}
\right\}Y(x;\bft),
\\
\frac{\partial Y(x;\bft)}{\partial t_i}&=
-\frac{\tilde{A}_i(\bft)}{x-t_i}Y(x;\bft) 
\quad (i=1,2,3), 
\end{align}
where the $2\times 2$ matrices $\tilde{A}_i(\bft)$ 
($i=1,2,3$) are defined as 
\begin{align}
\tilde{A}_i &\defeq 
-\begin{pmatrix} 
   (g_0^{-1})_{1i} \\
   (g_0^{-1})_{2i} 
     \end{pmatrix}
\begin{pmatrix} 
   (g_0)_{i1}(\beta_1 + 1) &
   (g_0)_{i2}(\beta_2 + 1) 
\end{pmatrix},
\end{align}
and satisfy the following relations:
\begin{equation}
\begin{aligned}
&\det\tilde{A}_i = 0,\quad
\mathrm{trace}\,\tilde{A}_i
=\mathrm{trace}\,A_i= -\alpha_i-1
\quad (i = 1, 2, 3),
\\
&-\tilde{A}_1 -\tilde{A}_2 -\tilde{A}_3 =
\begin{pmatrix}
\beta_1 + 1 &  0 \\ 0 & \beta_2 + 1
\end{pmatrix}.
\end{aligned}
\end{equation}
Thus we can show that the eigenvalues of $\tilde{A}_i$ 
are $0$ and $-\alpha_i -1$.

We now introduce new variables $\xi$ and $t$
as follows:
\begin{equation}
 \xi \defeq \frac{t_1-x}{t_1 - t_2}, 
\qquad
 t \defeq \frac{t_1 - t_3}{t_1 - t_2}.
\end{equation}
We finally obtain 
\begin{equation}
\frac{\partial Y}{\partial\xi}=
\left(\frac{\tilde{A}_1}{\xi}
+\frac{\tilde{A}_2}{\xi - 1}
+\frac{\tilde{A}_3}{\xi - t}\right)Y, \qquad
\frac{\partial Y}{\partial t}=
-\frac{\tilde{A}_3}{\xi - t}Y, 
\end{equation}
which is equivalent to \eqref{2by2Fuchsian}.
So we can obtain a Painlev\'e VI transcendent from 
the formula \eqref{solOfP6}. 
The parameters of the Painlev\'e VI in this case are 
given by 
\begin{equation}
\label{P6para}
\begin{aligned}
&\alpha = \frac{(\beta_1 - \beta_2 - 1)^2}{2},
\quad
 \beta = -\frac{(\alpha_1+1)^2}{2},\\
&\gamma = \frac{(\alpha_2+1)^2}{2},
\quad
 \delta = \frac{1-(\alpha_3+1)^2}{2}.
\end{aligned}
\end{equation}

\section{Action of the affine Weyl group $W(A_{n-1}^{(1)})$}
\label{sec:WeylGp}
In this section, we discuss two actions of the 
the affine Weyl group $W(A_{n-1}^{(1)})$, 
along the method developed in \cite{KK1,KK2}. 
Let $s_i$ $(1 \le i \le n-1)$ be the permutation $(i,i+1)$ 
and $S_i$ be the corresponding permutation matrix, 
\begin{equation}
 S_i \defeq E_{i,i+1} + E_{i+1,i} + \sum_{j \neq i,i+1} E_{jj}.
\end{equation}
We define the left-action $s_i^\mathrm{L}$ ($i=1, \dots, n-1$) by
applying $S_i$ to the initial data $g(0)$ from the left:
\begin{equation}
 s_i^\mathrm{L}g(\bft)
\defeq 
 \Psi_0(z; \bft) S_i g(0)
= S_i \Psi_0(z; s_i(\bft)) g(0)
= S_i g(s_i(\bft)),
\end{equation}
where the action of $s_i$ to the time-variables $\bft$ is 
defined as
\begin{equation}
s_i(\bft) \defeq 
(t_j^{s_i(1)}, \dots, t_j^{s_i(n)})_{j > 0}.
\end{equation}
Here we have used the relation $S_i \Lambda_j^a S_i^{-1}
= \Lambda_j^{s_i(a)}$ $(1\le a \le n, 1 \le i \le n-1,
j>0)$.

Applying the homogeneous Gauss decomposition \eqref{GaussDecomp} to 
$\Psi_0(z;t) S_i g(0)$, we define 
$s_i^\mathrm{L}g_{<0}(\bft)$ and $s_i^\mathrm{L}g_{\ge 0}(\bft)$:
\begin{equation}
\label{GaussLeftaction}
 s_i^\mathrm{L}g(\bft)
= \left\{ s_i^\mathrm{L}g_{<0}(\bft) \right\}^{-1}
        s_i^\mathrm{L}g_{\ge 0}(\bft).
\end{equation}
This decomposition induces an action of $s_j$ on the 
matrix elements of $g_{\ge 0}$ and $g_{<0}$.  
\begin{lemma}
\label{lemma_leftaction}
Assume that the Gauss decomposition 
\eqref{GaussLeftaction} exists. Then 
the action of $s_i$ $(i=1, \dots, n-1)$ on $g_{\ge 0}$ are
represented as the permutation of the $i$-th and 
the $(i+1)$-th raws: 
\begin{gather}
\label{LeftactionToqr0}
 (s_i^\mathrm{L}g_{\ge 0}(\bft))_{jk} 
= g_{\ge 0}(s_i(\bft))_{s_i(j) k}.
\end{gather}
\end{lemma}
\begin{proof}
The relation \eqref{GaussLeftaction} can
be written as follows:
\begin{equation}
\left\{s_i^\mathrm{L}g_{<0}(\bft)\right\}^{-1}
       s_i^\mathrm{L}g_{\ge 0}(\bft) = 
 S_i g(s_i(\bft)) = 
 S_i \left\{g_{<0}(s_i(\bft)) \right\}^{-1}
    g_{\ge 0}(s_i(\bft)).
\end{equation}
By multiplying $g_{<0}(s_i(\bft))$ from the left
and $g_{\ge 0}^{-1}(s_i(\bft))$ from the right, 
we have
\begin{equation}
\begin{aligned}
\lefteqn{
g_{<0}(s_i(\bft)) S_i \{ g_{<0}(s_i(\bft) \}^{-1}}
\qquad\\
&= \underbrace{g_{<0}(s_i(\bft)) 
   \{ s_i^\mathrm{L}g_{<0}(\bft) \}^{-1}}_{\in \mathfrak{g}_{<0}}
\cdot \underbrace{s_i^\mathrm{L}g_{\ge 0}(\bft) 
      \{ g_{\ge 0}(s_i(\bft)) \}^{-1}}_{\in \mathfrak{g}_{\ge 0}}.
\end{aligned}
\end{equation}
Due to the uniqueness of the Gauss decomposition 
\eqref{GaussLeftaction}, 
we obtain the left-action for $g_{<0}(\bft)$ and 
$g_{\ge 0}(\bft)$: 
\begin{align}
 &s_i^\mathrm{L}g_{<0}(\bft)
= \left( g_{<0}(s_i(\bft)) S_i \{g_{<0}(s_i(\bft)) \}^{-1} \right)_{<0}
 g_{<0}(s_i(\bft)),
\label{leftSi-}
\\
 &s_i^\mathrm{L}g_{\ge 0}(\bft) 
= \left( g_{<0}(s_i(\bft)) S_i \{g_{<0}(s_i(\bft)) \}^{-1} \right)_{\ge 0}
   g_{\ge 0}(s_i(\bft))
\label{leftSi+}
\end{align}
Notice that $S_i \in \mathfrak{g}_0 \subset
\mathfrak{g}_{\ge 0}$ holds
because we consider the homogeneous Gauss
decomposition. So the relation 
$
 \left( g_{<0}(s_i(\bft)) S_i \{ g_{<0}(s_i(\bft)) \}^{-1} \right)_{\ge 0}
= S_i
$ holds and thus we have
\begin{equation}
 s_a^\mathrm{L}g_{\ge 0}(\bft) 
= S_a g_{\ge 0}(s_a(\bft)).
\end{equation}
This leads to the desired results.
\end{proof}

Next we consider the symmetry of the similarity solution 
of the $\widehat{\mathfrak{gl}}_n$ hierarchy.
Under the similarity condition \eqref{simg},
we have 
\begin{equation}
\begin{aligned}
 s_i^\mathrm{L}g(z;\bft)
&= S_i \lambda^{D(\valpha)} 
   g(\lambda^{-1}z; s_i(\bft_\lambda)) \lambda^{D(\vbeta)}
\\
&= \lambda^{D(s_i(\valpha))} 
   s_i^\mathrm{L}g(\lambda^{-1}z; \bft_\lambda) 
  \lambda^{D(\vbeta)},
\end{aligned}
\end{equation}
where the action of $s_i$ to the parameters $\valpha$ is 
defined as
\begin{equation}
 s_i(\valpha) = (\alpha_{s_i(1)}, \dots, \alpha_{s_i(n)}).
\end{equation}
Then we see that the left-action of 
the Weyl group is realized as the permutation of 
the $i$-th and the $(i+1)$-th elements of 
the parameter $\valpha$ 
in the similarity condition \eqref{simg}.

The left-action of $s_i$ ($1\le i\le n-1$) induces 
a permutation of the coefficients in the
Schlesinger system \eqref{66_a}, \eqref{66_b}.
\begin{proposition}
\label{prop3}
The left-action of the 
Weyl group $s_i^\mathrm{L}$ ($1 \le i \le n-1$) induces
a permutation of the coefficient matrices of 
the equation \eqref{66_a} as follows:
\begin{equation}
 s_i^\mathrm{L} 
\left(\sum_{j=1}^n \frac{A_j}{x-t_j} \right)
= \sum_{j=1}^n \frac{A_{s_i(j)}}{x-t_{s_i(j)}}.
\end{equation}
\end{proposition}
\begin{proof}
By the definition of $A_j$ \eqref{def_Aa}, 
the $(k,l)$-element of $A_j$ is written as
\begin{equation}
 (A_j)_{kl} = -(g_0^{-1})_{kj} g_{jl} (\beta_l+1)
= -\frac{\Delta_{jk}(g_0)}{\det g_0} g_{jl} (\beta_l+1),
\end{equation}
where $\Delta_{jk}(g_0)$ is the $(j,k)$-cofactor of $g_0$.
By virtue of Lemma $\ref{lemma_leftaction}$, 
we can compute the action of $s_i^\mathrm{L}$ to 
$\Delta_{jk}(g_0)$, $\det g_0$ and $g_{jl}$ explicitly:
\begin{equation}
 s_i^\mathrm{L}(\Delta_{jk}(g_0)) = -\Delta_{s_i(j),k}(g_0),
\quad
 s_i^\mathrm{L}(\det g_0) = -\det g_0, 
\quad 
 s_i^\mathrm{L}(g_{jl}) = g_{s_i(j)l}. 
\end{equation}
We thus obtain 
\begin{equation}
 s_i^\mathrm{L}(A_j) = A_{s_i(j)},
\end{equation}
which completes the proof.
\end{proof}

If we focus on the $\widehat{\mathfrak{gl}}_3$-case, 
the left action of Weyl group
in Proposition \ref{prop3} gives transformations of the 
solutions of the Painlev\'e VI equation, which are given by 
\begin{alignat}{2}
s_1^\mathrm{L} :\;
&t \mapsto  1 - t,
&\quad
&y \mapsto 1 - y
\\
s_1^\mathrm{L}
s_2^\mathrm{L}
s_1^\mathrm{L} :\;
&t \mapsto \frac{t}{t - 1},
&\quad
&y \mapsto \frac{t - y}{t - 1}
\end{alignat}
These transformations of the Painlev\'e VI 
coincide with those included in \cite{Ok2}.

We then consider the action of $s_0\in W(A_{n-1}^{(1)})$. 
If we define a matrix $S_0$ as 
\begin{equation}
 S_0 \defeq zE_{n1} + z^{-1}E_{1n} + \sum_{j = 2}^{n-1}E_{jj}, 
\end{equation}
the left-action $s_0^\mathrm{L}$ can be 
constructed in the same manner. 
Furthermore, we consider the translations in $W(A_{n-1}^{(1)})$, 
i.e., 
\begin{equation}
\begin{aligned}
 T_i &= S_{i+1} S_0 S_{i+1} S_i\\
&= zE_{ii} + z^{-1}E_{i+1,i+1} + \sum_{j \neq i, i+1}E_{jj},
\end{aligned}
\end{equation}
for $i=1, \dots, n-1$, where we define $S_n=S_1$.
The action of $T_i$ ($i=1, \dots, n-1$) to the similarity solution 
makes a change in the parameters $\vec{\alpha}$, which 
is equivalent to the Schlesinger transformation.

We can also define the right-action of $s_i\in A_{n-1}$ 
($i=1,\dots,n-1$) by using the action of $S_i$ to the 
initial data $g(0)$ \cite{KK1}. 
This action causes the permutation of the $i$-th and the 
$(i+1)$-th 
columns of $g_{\ge 0}(\bft)$:
\begin{equation}
 (s_i^\mathrm{R}g_{\ge 0}(\bft))_{jk} 
= g_{\ge 0}(s_i(\bft))_{j s_i(k)}.
\end{equation}
Furthermore, the right-action of $s_i$ 
($i=1,\dots,n-1$) to the
similarity solution causes the permutation of 
the $i$-th and the $(i+1)$-th elements of 
the parameters $\vec{\beta}=(\beta_1,\dots,\beta_n)$.

\section{Concluding remarks}
\label{sec:ConcludingRemarks}
In this paper, we have established the method for 
obtaining the Painlev\'e VI from the 
$\widehat{\mathfrak{gl}}_3$ hierarchy. 
In the $\widehat{\mathfrak{gl}}_n$-case, 
we can obtain the $2 \times 2$ Schlesinger system 
that is associated with the Garnier system \cite{Ok} from 
\eqref{66_a}, \eqref{66_b} by 
setting $\beta_3 = \beta_4 = \cdots = \beta_n = -1$.
Hence our method is valid also for the Garnier system. 

We also consider the two actions of the affine Weyl group 
$W(A_{n-1}^{(1)})$ to the $\widehat{\mathfrak{gl}}_n$ hierarchy 
along the method used in \cite{KK1,KK2}. 
The left-action is realized as the permutations of the 
parameters $\vbeta$. 
In the case of $n=3$, these actions corresponds to 
symmetry properties of the Painlev\'e VI equation.
However, the symmetry discussed in this paper is 
a subset of the full-symmetry of the Painlev\'e VI, 
which is isomorphic to $W(F_4^{(1)})$ \cite{Ok2}. 
It may be worthwhile to investigate 
the meaning of other elements in $W(F_4^{(1)})$, 
in the context of the $\widehat{\mathfrak{gl}}_3$ hierarchy.

\section*{Acknowledgments}
The authors would like to thank Professors 
K.~Hasegawa, T.~Ikeda, A.V.~Kitaev, G.~Kuroki,
Y.~Ohyama, K.~Okamoto, H.~Sakai and K.~Takasaki
for their interests and discussions. 
The first author is partially supported by 
the Grant-in-Aid for Scientific Research (No.~16740100) from 
the Ministry of Education, Culture, Sports, Science and Technology.
The second author is partially supported by 
the 21st Century COE program of Tohoku University:
Exploring New Science by Bridging Particle-Matter Hierarchy.


\begin{thebibliography}{9999}
\bibitem[AvdL]{AV}
Aratyn, H. and van de Leur, J.:
Integrable  structure behind WDVV equations,
\textit{Theoret.~Math.~Phys.} \textbf{134} (2003), 14--26.

\bibitem[BtK1]{BtK1} 
Bergvelt, M.~J. and ten Kroode, A.~P.~E.: 
$\tau$ functions and zero curvature equations of Toda-AKNS type,
\textit{J.~Math.~Phys.} \textbf{29} (1988), 1308--1320.

\bibitem[BtK2]{BtK2} 
Bergvelt, M.~J. and ten Kroode, A.~P.~E.: 
Partitions, vertex operator constructions 
and multi-component KP equations,
\textit{Pacific J.~Math.} \textbf{171} (1995), 23--88.

\bibitem[BdGHM]{BdGHM}
Burroughs, N.~J., de Groot, M.~F., 
Hollowood, T.~J. and Miramontes, J.~L.:
Generalized Drinfeld-Sokolov hierarchies II: 
the Hamiltonian structures, 
\textit{Commun.~Math.~Phys.} \textbf{153} (1993), 187--215.

\bibitem[DS]{DS} 
Drinfel'd, V.~G. and Sokolov, V.~V.:  
Lie algebras and equations of Korteweg-de Vries type,
\textit{J.~Sov.~Math.} \textbf{30} (1985), 1975--2036.

\bibitem[Du]{Dub} 
Dubrovin, B.: 
Geometry of $2$d topological field theories,
integrable systems and quantum groups,
\textit{Springer Lecture Notes in Math.} \textbf{1620},
Springer, Berlin (1996), 120--348.

\bibitem[FY]{FY} 
Fokas, A.~S. and Yortsos, Y.~C.: 
The transformation properties of the
sixth Painlev\'e equation and one-parameter
families of solutions, 
\textit{Lett.~Nuovo Cimento} \textbf{30} (1981), 539--544.

\bibitem[dGHM]{gds1} 
de Groot, M.~F., Hollowood, T.~J. and Miramontes, J.~L.: 
Generalized Drinfel'd-Sokolov hierarchies,
\textit{Commun.~Math.~Phys.} \textbf{145} (1992), 57--84.

\bibitem[H]{Har} 
Harnad, J.: 
Dual isomonodromic deformations
and moment maps to loop algebras, 
\textit{Commun.~Math.~Phys.} \textbf{166} (1994), 337--365.

\bibitem[JM]{JM2}
Jimbo, M. and Miwa, T.: 
Monodromy preserving deformation of linear
ordinary differential equations with rational coefficients. II,
\textit{Physica} \textbf{D2} (1981), 407--448.

\bibitem[K]{Kac}
Kac, V.~G.: 
\textit{Infinite dimensional Lie algebras}, third edition.
Cambridge University Press, 1990.

\bibitem[KvdL]{KvdL}
Kac, V.~G. and van de Leur, J.: 
The $n$-component KP hierarchy
and representation theory, in 
\textit{Important developments
in soliton theory}, Springer, Berlin (1993), 302--343. 

\bibitem[KK1]{KK1}
Kakei, S. and Kikuchi, T.: 
Affine Lie group approach to a derivative
nonlinear Schr\"odinger equation and
its similarity reduction,
\textit{Int.~Math.~Res.~Not.} {\bf 78} (2004),
4181--4209.

\bibitem[KK2]{KK2}
Kakei, S. and Kikuchi, T.: 
Solutions of a derivative nonlinear Schr\"odinger hierarchy 
and its similarity reduction, 
\textit{Glasgow Math.~J} \textbf{47}, Issue A (2005), 99--107.

\bibitem[KIK]{KIK}
Kikuchi, T., Ikeda, T. and Kakei, S.: 
Similarity reduction of the modified Yajima-Oikawa equation,
\textit{J.~Phys. A: Math.~Gen.} \textbf{36} (2003), 11465--11480.

\bibitem[K]{Kit} 
Kitaev, A.~V.: 
On similarity reductions of the three-wave resonant
system to the Painlev\'e equations, 
\textit{J.~Phys. A: Math.~Gen.} \textbf{23} (1990), 3543--3553.

\bibitem[M]{Maz} 
M.~Mazzocco, 
Painlev\'e sixth equation as
isomonodromic deformations equation
of an irregular system,
In \textit{The Kowalevski property}: 
CRM Proceedings and Lecture Notes
\textbf{32} (2002), 219--238.

\bibitem[N]{ICM} 
M.~Noumi,
Affine Weyl group approach to Painlev\'e equations,
in {\it Proceedings of the International Congress of
Mathematicians} (Tatsien, L., ed.), Vol.~III, 
World Scientific (2002), 497--509.

\bibitem[NY]{NYA} 
Noumi, M. and Yamada, Y.: 
Higher order Painlev\'e equations of type $A^{(1)}_l$,
\textit{Funkcial. Ekvac.} {\bf 41} (1998), 483--503.

\bibitem[O1]{Ok}
Okamoto, K.: 
Isomonodromic deformation and Painlev\'e equations,
and the Garnier system, 
\textit{J. Fac.~Sci.~Univ.~Tokyo}, 
Sect.~IA, Math. \textbf{33} (1986), 575--618.

\bibitem[O2]{Ok2}
Okamoto, K.: 
Studies on the Painlev\'e equations I.
Sixth Painlev\'e equation $\mathrm{P_{VI}}$,
\textit{Ann.~Mat.~Pura Appl.} (4) \textbf{146} (1987), 337--381.

\bibitem[UT]{UT}
Ueno, K. and Takasaki, K.: 
Toda lattice hierarchy, 
\textit{Adv.~Stud.~in Pure Math.} \textbf{4} (1984), 1-95.

\bibitem[W]{Wil}
Wilson, G.: 
The $\tau$-functions of the $\mathfrak{g}$AKNS equations,
in {\it Verdier memorial conference on integrable systems}
(Babelon, O., Cartier, P. and Kosmann-Schwarzbach, Y., eds.),
Birkhauser (1993), 131--145.
\end{thebibliography}
\end{document}